\documentclass
[superscriptaddress,secnumarabic,amssymb,amsmath,nobibnotes,aps,prd,showkeys,nofootinbib,nopacsnumber,onecolumn,10pt]{revtex4}%
\usepackage{graphicx}
\usepackage{bm}
\usepackage{xcolor}
\usepackage{amsmath}
\usepackage{amsfonts}
\usepackage[english]{babel}
\usepackage{amssymb}
\usepackage[utf8]{inputenc}%
\providecommand{\U}[1]{\protect\rule{.1in}{.1in}}

\newcommand{\be}{\begin{equation}}
\newcommand{\ee}{\end{equation}}

\newcommand{\mincir}{\raise
-3.truept\hbox{\rlap{\hbox{$\sim$}}\raise4.truept\hbox{$<$}\ }}
\newcommand{\magcir}{\raise
-3.truept\hbox{\rlap{\hbox{$\sim$}}\raise4.truept\hbox{$>$}\ }}

\begin{document}
\title{{National Mapping and Testing of Astronomical Sites in Ethiopia (NMTASE)}}
\author{\textbf{SSGI Collbaration}\footnote{SSGI:Space Science and Geo-Spatial Institute}:  Shambel Sahlu}
\email{shambel.sahlu@nithecs.ac.za}
\affiliation{Centre for Space Research, North-West University, Potchefstroom 2520, South Africa}
\affiliation{Entoto Observatory and Research Center, Space Science and Geo-Spatial Institute, Addis Ababa, Ethiopia}
\affiliation{National Institute for Theoretical and Computational Sciences (NITheCS), South Africa.}
\author{Nebiyu Suleyman}  
\affiliation{Ethio Telecom, Addis Ababa, Ethiopia.}
\author{Gemechu M. Kumssa}
\affiliation{Entoto Observatory and Research Center, Space Science and Geo-Spatial Institute, Addis Ababa, Ethiopia}
\author{Solomon T. Belay}
\affiliation{Entoto Observatory and Research Center, Space Science and Geo-Spatial Institute, Addis Ababa, Ethiopia}
\author{Feraol F. Dirirsa }
\affiliation{Entoto Observatory and Research Center, Space Science and Geo-Spatial Institute, Addis Ababa, Ethiopia}
\author{Esubalew Mulugeta}
\affiliation{Entoto Observatory and Research Center, Space Science and Geo-Spatial Institute, Addis Ababa, Ethiopia}
\author{Melesse Getnet}
\affiliation{Entoto Observatory and Research Center, Space Science and Geo-Spatial Institute, Addis Ababa, Ethiopia}
\author{ Mirjana Povic}
\affiliation{Entoto Observatory and Research Center, Space Science and Geo-Spatial Institute, Addis Ababa, Ethiopia}
\author{Seblu H. Negu}
\affiliation{Entoto Observatory and Research Center, Space Science and Geo-Spatial Institute, Addis Ababa, Ethiopia}
\author{Betelehem Bilata}
\affiliation{Entoto Observatory and Research Center, Space Science and Geo-Spatial Institute, Addis Ababa, Ethiopia}
\author{ Jerusalem Tamrat}
\affiliation{Entoto Observatory and Research Center, Space Science and Geo-Spatial Institute, Addis Ababa, Ethiopia}

\begin{abstract}
This work aims to choose potential astronomical sites that can be candidates for a new astronomical optical observatory in Ethiopia, in addition to the Entoto Observatory and Lalibela sites. For our primary investigation, the six basic criteria, namely the altitude of the mountains, artificial light pollution, cloud coverage, humidity, wind speed, and wind direction, were taken into account.  Consequently, using the multi-criteria statistical Decision analysis (MCDSA) techniques, 21 high-potential places are selected and presented for further investigation out of 367 mountains. Among these 21 candidates, three sites, Bauhit, Meseraia, and T’at’a  are the most suitable places for optical astronomy in Ethiopia. Those selected mountains are mapped and presented to study the future of the astronomical seeing effect. This study may contribute to the protection of those potential astronomical sites and their dark skies and the development of astrotourism for the sustainable development of modern astronomy in Ethiopia and in the East African region. 
\end{abstract}
\date{\today}
\maketitle

\section{Introduction} 
Ethiopia holds a highly unique and stratagic positions in the global Astronomy ladscape, referred to as the roof of Africa. It has an advantageous location near the equator, which can have access to see the north and south sky at the same time and its expansive plateaus, towering up to 4,550 metres above sea level, combined
with its prevailing dry weather, establish it as an exceptional region for premier astronomical observation sites on a global scale. Ethiopia has two sites ongoing, namely Entoto Mountain and the Lalibela site. However, due to the expansion of Addis Ababa (the Capital of Ethiopia) and Sululta cities, Entoto Observatory has been extremely affected by artificial light pollution. In the current project of searching for tests of astronomical sites under the Ethiopian Space Science and Geospatial Institute, we highlighted the importance of reserving the best astronomical pace for an optical observatory in addition to the Entoto observatory and Lalibela sites by setting 11 standard criteria \cite{aksaker2020global,kebede2002astronomy,vernin1995measuring,guessoum2014preliminary,abdelaziz2017search} and employing the multi-criteria decision analysis technique (MCDA) and detailed statistical analysis.
\\
\\
In the meantime, we have successfully identified 21 promising locations as preliminary results based on six fundamental criteria: altitude, cloud cover, city light pollution, humidity, wind speed and direction, and slope.  These carefully selected sites capitalise on Ethiopia's unique
geographical features, including its equatorial proximity, elevated plateaus, and favourable weather conditions, while adhering to
rigorous astronomical site evaluation standards. In this work, we also take into account multiple factors, such as atmospheric conditions, weather patterns, dark skies, altitude, and geographic positioning, to ensure that the selected mountains have good sustainability for the observatory establishment. This research project combines MCDSA, utilising eleven key criteria to assess potential optical observatory sites. Particularly the three site places: Bauhit, Meseraia and T’at’a  are  very suitable places for optical astronomy possibly attract the golobe astronomy attaraction.  Reliable data are collected from primary and secondary sources, including meteorological satellite data, on-site measurements, and astronomical surveys. NASA's comprehensive and user-friendly data (\url{https://power.larc.nasa.gov/dataaccess-viewer/}) has been specifically chosen as a valuable resource for the project due to its long-term records and diverse parameters. In the future, this work aims to improve the site selection process by incorporating additional parameters, such as seismic vulnerability, aerosol optical concentration, availability of
infrastructure, diurnal temperature, and perceptible water vapour (PWV). Astronomical seeing \cite{vernin1995measuring,tillayev2021astronomical}, which is the most important assessment of the astronomical observatory, will be implemented for each selected mountain in the next phase of the work. The significance of these chosen areas for the astronomy community is profound, offering valuable resources for research and presenting enticing prospects for astronomical tourism in addition to sustainable scientific development.  
\\
\\
\section{Data and data sources}
All data that we have used for this work have been collected from a variety of sources\footnote{ Data sources: NASA \url{https://power.larc.nasa.gov/data-access-viewer/}, FAO \url{https://www.fao.org/aquastat/en/geospatial-information/climate-information}, Weather Spark  \url{https://weatherspark.com/}, and SRTM \url{https://www.earthdata.nasa.gov/sensors/srtm}) and NMA \url{http://www.ethiomet.gov.et/}} including astronomical surveys, satellite data, meteorological data, and on-site observations. Both primary and secondary data sources were used in the research; the data sources for each criterion are shown in Table \ref{data}. Weather trends at the possible location are tracked and predicted using meteorological satellite data, while on-site observations yield information about infrastructure and light pollution.
{\small
\begin{table}[h!]
    \centering
   \begin{tabular}{ |c|c|c|c| } 
 \hline
 No & Criteria  & Parameters&Data Source   \\ 
 \hline
 1& Altitude &Elevations above see level & SRTM, DEM, Weather Spark\\
 \hline
 2& Night Sky Brightness& Artificial Light Pollution & NASA\\ 
 \hline
 3 & Photometric Night fraction  & Cloud & NASA, FAO, Weather Spark\\ 
 \hline
 4&Relative Humidity&Relative Humidity & NASA, FAO, Weather Spark, EMI\\
 \hline
5 &Wind speed and direction &Wind speed and direction &  NASA, FAO, EMA\\
 \hline
\end{tabular}
    \caption{Data sources for the criteria}
    \label{data}
\end{table}
}
\\
\\
After collecting the necessary data from the above sources, we evaluated and compared the data quality and quantity. After assessing the accuracy and relevance of the data, we used the data from NASA( \url{https://power.larc.nasa.gov/data-access-viewer/}). Since the site provides reliable information, location-based data, comprehensive coverage, long-term records, a variety of parameters, and also it is a user-friendly interface. 
\section{Results and discussion}
\begin{figure}[!ht]
\includegraphics[scale=0.8]{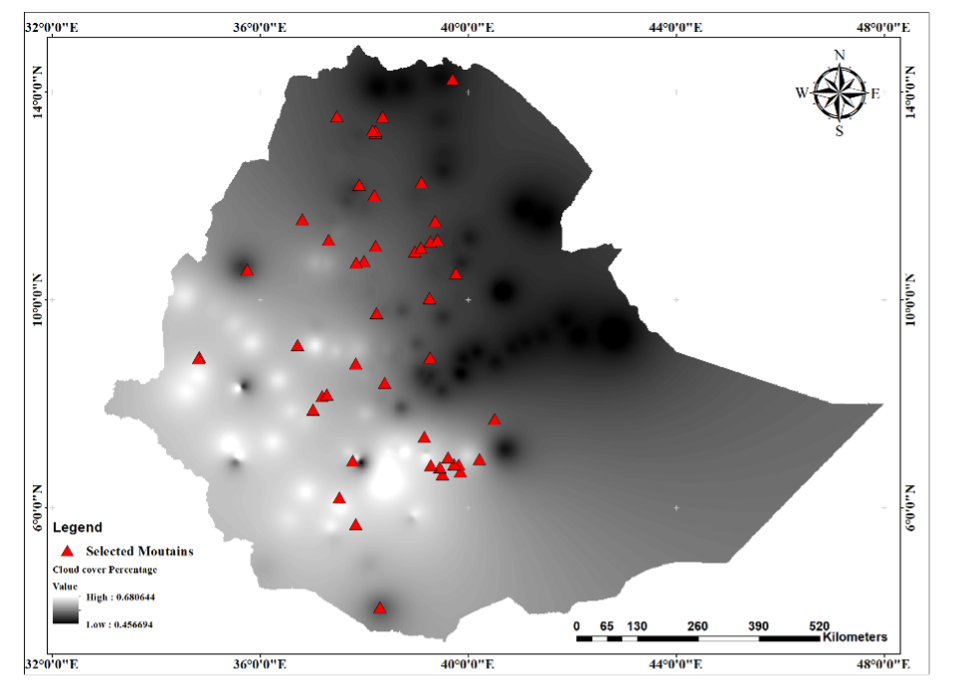}
\includegraphics[scale=0.4]{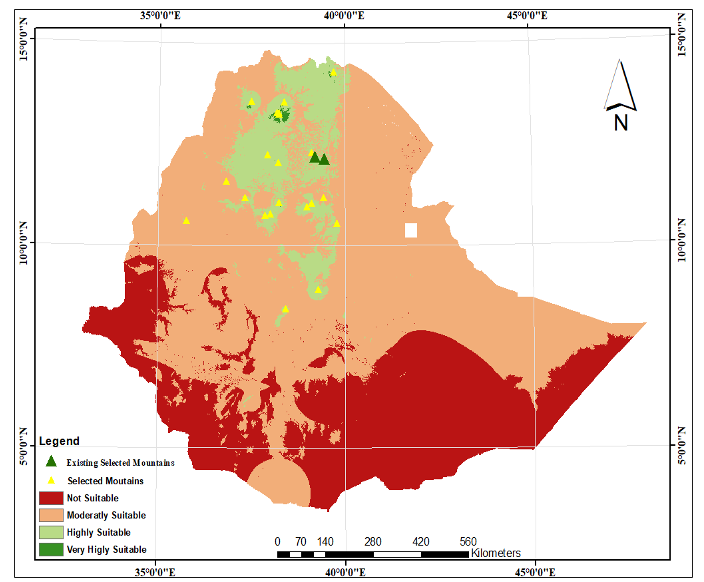}
\caption{Left Panel: 46 Selected mountains based on cloud. Right Panel: 21 selected mountains after we consider the six different criteria.   }
\label{fig:1}
\end{figure}

Following the six selection criteria mentioned earlier, we have set the mountain altitude as the primary criterion for selecting possible site locations and identified 367 mountains in Ethiopia that were labeled as being more than 2500 m above sea level. Then, next, we considered light pollution as a secondary criterion to proceed with other site selection methods.  Because, based on the recent IAU report 29 meeting, light pollution is becoming a serious problem, especially for the Astronomy Society. This problem comes from city expansion, the construction of infrastructure, and a lack of awareness to preserve and protect the astronomical places. For instance, with Europe, North America most of Asia are dominated by artificial light pollution, Africa, some parts of South America, and Australia have relatively less artificial light pollution \cite{falchi2016new} because of a lack of infrastructure development. We take into account Addis Ababa as a medium-sized city \cite{vernin1986astronomical} and used it as a reference to categorize all Ethiopian cities as Medium (I), Regional City (II), Zonal City (III), District (IV), Kebele (V), and Village (VI) by comparing their size with Addis Ababa and calculating the distance between the mountain (site places) and the nearby city using the Longitude and Altitude differences, and setting the following criteria for further investigations depend on the light pollution amounts, see Table \ref{qq}.
\textcolor{red}{
\begin{table}[h!]
    \centering
   \begin{tabular}{ |c|c|c| } 
 \hline
 Cities Category & City Type & Selected mountains   \\ 
 &&based distance from the city\\
 \hline
 Capital City & I& above 50 km \\ 
 \hline
 Regional City & II & above 35 km \\ 
 \hline
 Zonal City &III& above 25 km\\
 \hline
 Woreda City &IV& above 17km\\
 \hline
 Kebele City&V&above 10km\\
 \hline
 Village &VI&above 5km\\
 \hline
\end{tabular}
    \caption{Ethiopian cities categorized based on their light-pollution amount}
    \label{qq}
\end{table}
}
Based on the light pollution criteria, we found 228 mountains from 367 selected mountains throughout the country.  
\\
\\
Thirdly, one of the most important factors in choosing a possible location for an observatory for astronomy is cloudiness. Tens of kilometers away from the light source, light may be visible since clouds naturally reflect light. There is a definite relationship between cloud cover and height and night sky brightness. The brightness of the night sky can therefore vary greatly, even while light output is constant. As presented in \cite{Aksaker2020}, the astronomical site is preferred 70\% (mean 110 clear nights for observation) is cloudy throughout the year.  In our project, we consider 214 mountains which have a range of 45 - 68 \% cloudy mean of 201 to 116 clear nights throughout the year, respectively. At this stage, we selected 46 astronomical site places from 228 mountains and mapped them in the Right panel of Fig. \ref{fig:1}. 
\\
\\
Finally, the humidity \cite{vernin1986astronomical}, wind speed, and direction \cite{garcia2009adaptive} are considered as the fourth, fifth, and sixth selection criteria, respectively, that are crucial in choosing the potential places from the above 46 astronomical observatory sites. Since high humidity can cause atmospheric turbulence, blurring telescope images, and reducing clear observation nights due to cloud formation. Ideal sites prioritize humidity levels between 30\% to 70\% to minimize turbulence and cloud interference, and based on this, the numbers of mountains are reduced to 32 based on the humidity of the mountains.  Additionally, strong winds can disrupt telescope movement and induce atmospheric turbulence, affecting celestial object tracking. Optimal observatory sites maintain low wind speeds $<10m/s$  with consistent directions, complemented by low humidity levels. Based on the assessments of the wind speed and direction combined with MCDA, the numbers of mountains are reduced from 32 to 21, which are mapped in the Left Panel of Fig. \ref{fig:1}. 
\\
\\
In general, in this work, 21 suitable astronomical palaces together with the two existing sites (Entoto Observatory and Lalibela) are presented, with their name in the Table. \ref{}.  As mentioned earlier, we only consider the above six essential criteria from the eleven standard criteria.   Including the astronomical seeing, other important criteria, namely: Seismic variability, precipitable water vapor, seismic risk, aerosol concentration, and infrastructural accessibility, will be considered in the next phase of the work for future investigation. 

\begin{table}{}
\begin {tabular}{ |c|c|c|c|c|c|c|c|c| }
\hline
No & Mountain & Altitude & Cloud \% & Light & relative \% & Wind  & Suitability \\
&Name & in m &Amount& pollution&humidity& speed in m/s& index\\
\hline
1 & Abieri  & 3641 & 47.14943 & very very small &
55.3946341 & 3.40597561 & Highly Suitable&\\\hline
2 & Abune Yosef  & 4050 & 52.60379 & very very small &
64.80164634 & 3.676341463 & Highly Suitable & Existing Site\\
\hline
3 & Abuye Meda  & 3207 &
59.6084 & very very small & 57.9674187 & 3.00800813 &Highly Suitable &Existing Site\\
\hline
4&Amba Farit  & 3741 & 55.23434 &very very
small & 52.90577236 & 2.708313008 & Highly Suitable &\\
\hline
5 & Awrimo  & 2896 & 47.69651 & very very
small & 45.60067073 & 4.752184959 & Highly Suitable&\\
\hline
6 & Bauhit  & 4343 & 47.14943 & very
very small & 58.11067073 & 3.225457317 & Very Highly Suitable&\\
\hline
7 & Boccan  & 2570 & 51.86934 &
very very small & 62.7650813 & 3.977743903 & Highly Suitable &\\
\hline
8 & Carni  & 3335 & 55.23434 & very very small
& 57.99430894 & 2.721178862 & Highly Suitable &\\
\hline
9 & Chef  & 2602 & 54.5314 & very
very small & 57.99430894 & 2.721178862 & Highly Suitable&\\
\hline
10&Ch'ok'e  & 4031&63.88417 &very very
small&63.31979675&2.850050813&Highly Suitable&\\
\hline
11 & Culan  & 2522 & 54.23651 & very
very small & 68.40788618 & 3.018394309 & Highly Suitable&\\
\hline
12 & Gurage  & 3539 & 54.47015 &
very very small & 68.45670732 & 4.189705285 & Highly Suitable&\\
\hline
13 & Ioll  & 4202 & 56.16463 & very very small
& 67.04623984 & 3.508953252 & Highly Suitable&\\
\hline
14 & Israel Amba  & 2974 & 52.60379 &
very very small & 61.29254065 & 4.039867886 & Highly Suitable&\\
\hline
15 & Lieu  & 3074 & 53.87507
& very very small & 66.71802846 & 2.883434959 & Highly Suitable &\\
\hline
16 & Melza  & 2722 & 45.13963 &
very very small & 57.98142276 & 3.280792683 & Highly Suitable&\\
\hline
17 & Meseraia  & 4214 & 47.14943 & very
very small & 55.39463415 & 3.40597561 & Very Highly Suitable&\\
\hline
18 & Sena  & 2590 & 53.23018 & very very
small & 66.53979675 & 3.336626016 & Highly Suitable&\\
\hline
19 & Socona  & 2723 & 52.60379 & very very small
& 64.80164634 & 3.676341463 & Highly Suitable &\\
\hline
20 & Tagle Gheorghis  & 2600 & 47.2655 & very
very small & 52.25376016 & 2.951443089 & Highly Suitable&\\
\hline
21 & T'at'a  & 4058 & 47.14943 & very
very small & 55.39463415 & 3.40597561 & Very Highly Suitable&\\
\hline
22 & Uenfit  & 2830 & 52.17844
& very very small & 57.98142276 & 3.280792683 & Highly Suitable&\\
\hline
23 & Uwa  & 3305 & 56.16463 & very very
small & 67.04623984 & 3.508953252 & Highly Suitable&\\
\hline
\end {tabular}
\caption{ Based on the six essential criteria, 21 selected potential site places  are shown in this table together with two existing site places (Entoto and Lalibela)}
\label{}
\end{table}
\section{Conculusions}
Conducted under the auspices of the Space Science and Geospatial Institute, this project aims to strategically evaluate and validate potential astronomical sites across the nation. Utilizing a multi-criteria decision analysis based on six essential geographical and atmospheric factors, the study successfully narrowed an initial selection of 367 Ethiopian mountains to 21 high-potential locations. Among these, Bauhit, Meseraia, and T’at’a were identified as the most viable sites to sustainably advance Ethiopia's capabilities in optical astronomy. To ensure the long-term feasibility of these locations, the upcoming phase of the project will integrate three critical dimensions: physical site assessments to review infrastructure readiness in coordination with local and regional authorities, seismic datasets to evaluate regional stability, and detailed observational seeing measurements.

\section*{Acknowledgements}
The authors express their gratitude to the Space Science and Geospatial Institute (SSGI) for providing the financial support that enabled this research. We also highly appreciate the preliminary findings of Nebiyu Suleyman, who initially identified 108 site configurations across 191 mountains. Building upon those initial suggestions, this study expanded the screening framework to encompass 367 mountains, ultimately narrowing the selection to the 21 high-potential candidate sites presented herein.
\bibliographystyle{iopart-num}
\bibliography{biblio}
\end{document}